# Sparking mashups to form multifunctional alloy nanoparticles

Jicheng Feng[1]†*, Dong Chen[2]†, Peter V. Pikhitsa[1], Yoon-ho Jung[3], Jun Yang[2]*, Mansoo Choi[1,3]*

[1]Global Frontier Centre for Multiscale Energy Systems, Seoul National University, Seoul 08826, Republic of Korea

[2]State Key Laboratory of Multiphase Complex Systems, Institute of Process Engineering, Chinese Academy of Sciences, Beijing 100190, China

[3]Department of Mechanical and Aerospace Engineering, Seoul National University, Seoul 08826, Republic of Korea

†These authors contributed equally to this work.

*All correspondence should be addressed to jic.feng@gmail.com (J.F.), jyang@ipe.ac.cn (J.Y.), mchoi@snu.ac.kr (M.C.)

Synthesizing unconventional alloys remains challenging owing to seamless interactions between kinetics and thermodynamics. High entropy alloys (HEAs), for example, draw a fundamentally new concept to enable exploring unknown regions in phase diagrams. The exploration, however, is hindered by traditional metallurgies based on liquid–solid transformation. Vapor–solid transformation that is permissible on pressure–temperature phase diagrams, offers the most kinetically efficient pathway to form any desired alloy (e.g., HEA). Here, we report that a technique called "sparking mashups", which involves a rapidly quenched vapor source and induces unrestricted mixing for alloying 55 distinct types of ultrasmall nanoparticles (NPs) with controllable compositions. Unlike the precursor feed in wet chemistry, a microseconds-long oscillatory spark controls the vapour composition, which is eventually retained in the alloy NPs. The resulting NPs range from binary to HEAs with marked thermal stability at room temperature. We show that a nanosize-effect ensures such thermal stability and mimics the role of mixing entropy in HEAs. This discovery contradicts the traditional "smaller is less stable" view while enabling the elemental combinations that have never been alloyed to date. We even break the miscibility limits by mixing bulk-immiscible systems in alloy NPs. As powerful examples, we demonstrate the alloy NPs as both high-performance fuel-cell catalysts and building blocks for three-dimensional (3D) nanoprinting to construct HEA nanostructure arrays of various architectures and compositions. Our results form the basis of new rules for guiding HEA-NP synthesis and advancing catalysis and 3D printing to new frontiers.

Unconventional alloy synthesis is a complex process because of the seamless interactions between thermodynamics and kinetics and the excessive dependence on the production conditions (e.g., applied techniques, used precursors). High-entropy alloys (HEAs) concept introduces a fundamentally novel strategy for exploring unknown regions in phase diagrams[1]. The phase stability of HEAs was originally thought to relate only to maximizing entropy[1,2], but driving forces often overcome entropy contribution to form secondary phases[3,4]. Together with phase selection rules[4–6], these driving forces stipulate the involvement of other influential factors, such as the mixing enthalpy ($\Delta H_{mix}$). Particularly, bulk-immiscible systems ($\Delta H_{mix}>0$)[7] establish energy barriers against alloying in thermodynamic equilibrium (mixing Gibbs free energy $\Delta G_{mix} = \Delta H_{mix} - T\Delta S_{mix}$ with mixing entropy $\Delta S_{mix}$ at temperature $T$). A non-equilibrium process, however, can kinetically trap a random alloy, in which stability is ensured by nanoscale confinement[8]. Even nanograined unary metals exhibit notable thermal stability[9]. Further, HEAs reported thus far have been restricted to a palette of similar atoms, and groupings of atoms with vastly different chemical and physical properties have rarely been reported[10]. The main hindrances to such HEAs include limited theoretical access and a lack of synthesis techniques that enforce the unrestricted mixing of any elements of interest. Traditional metallurgies for making HEAs have not yet successfully fabricated the full range of compositions permitted in theoretical phase diagrams, thus necessitating the development of new synthetic strategies.

Nanoparticles (NPs) have become integral building blocks in material design[11], and further mixing multiple components within single NPs paves the way to new materials with predetermined functionalities[2]. Wet-chemistry synthesis delivers remarkable controllability in NP size and morphology, but extending this approach to multicomponent alloy (MA) NPs imposes tight constraints (e.g., immiscibility, phase selection, > 3 elements) and undue complications[2]. In addition, NP quality seems inseparable from substrate properties[2,12,13]. Complete alloying is typically prevented by wide gaps between the reduction potentials[14,15], although this difference has been revealed to be a practicable way to hollow out NPs[16]. Electrostatic adsorption was exploited to synthesize ultrasmall bimetallic alloy NPs, as evidenced only by the so-called speckling effect[12]. An analogous method reduced sequentially adsorbed heterometallic precursors to synthesize bimetallic NPs, but pairs of components that are immiscible in the bulk exhibited sub-nanometre intraparticle phase segregation[13]. The carbothermal shock method succeeded in fabricating various MA-NPs, though this method requires defective carbon nanofibers as supports[2]. Interestingly, in one study, small MA-NPs exhibited superior stability despite deep annealing[17], whereas other works showed that slightly larger NPs held precisely defined interfaces[18,19]. Unlike in MA-NPs, such heterostructure NPs appear to have limited synergistic benefits. Targeted applications require the HEA-NPs to have precisely tailored compositions and sizes, but a fundamental understanding to achieve such requirements remains lacking[20]. Further, the scalability of HEA-NP fabrication remains a formidable obstacle, mainly in connection with slow kinetics in liquid–solid transformation and minute NP loadings on a mass support.

Here, we propose that a vapour-source technique called "sparking mashups" provides an unrestricted mixing environment for alloying 55 distinct types (table S1) of ultrasmall NPs (<5 nm) with controllable compositions. The synthesized NPs range from binary alloys to HEAs, including bulk-immiscible elements (table S1, fig. S1) and element combinations that have never been alloyed so far (table S2), to the best of our knowledge. In sparking mashups, two or more constituent materials are vaporized by an oscillatory spark with a duration of microseconds, followed by ballistic transport and intermixing to form alloy NPs. To describe this sparking and mixing process, we coined the term "sparking mashups". The resulting NPs are then carried by a high-purity gas to a substrate. In wet chemistry, the feeding ratio of precursors is changed to control the composition, whereas in this method, an oscillating spark controls the vapour composition, which is retained in the final alloy NPs. Alloy NP formation follows the most efficient kinetic pathway that leads the metal vapours to a thermodynamically stable state. Considering that metallic bonding is electronic in nature and underlies quantum physical effects, our method enables alloying any dissimilar metals in NPs. Below we discuss alloy NPs composed of such bulk-immiscible systems. Our results confirm that intermetallics are indeed kinetically suppressed, thus favouring random alloys. We show that the nanosize effect mimics the role of entropy in HEAs to stabilize MA-NPs, thereby extending the HEA concept to new frontiers. As powerful examples, we first demonstrate their promising catalytic activities in three fuel-cell reactions. Next, as building blocks for three-dimensional (3D) printing, the alloy NPs were printed into 3D nanostructure arrays of various alloys and architectures. In this work, we not only paired the predicted and experimental synthesis of MA-NPs but also demonstrated flexible means and substrates for capturing such NPs. Therefore, our general approach discloses a powerful roadmap to new and uncharted territory, thus advancing the fields of HEAs, catalysis, and 3D printing.

## Vapour–solid transformation into any desired alloy

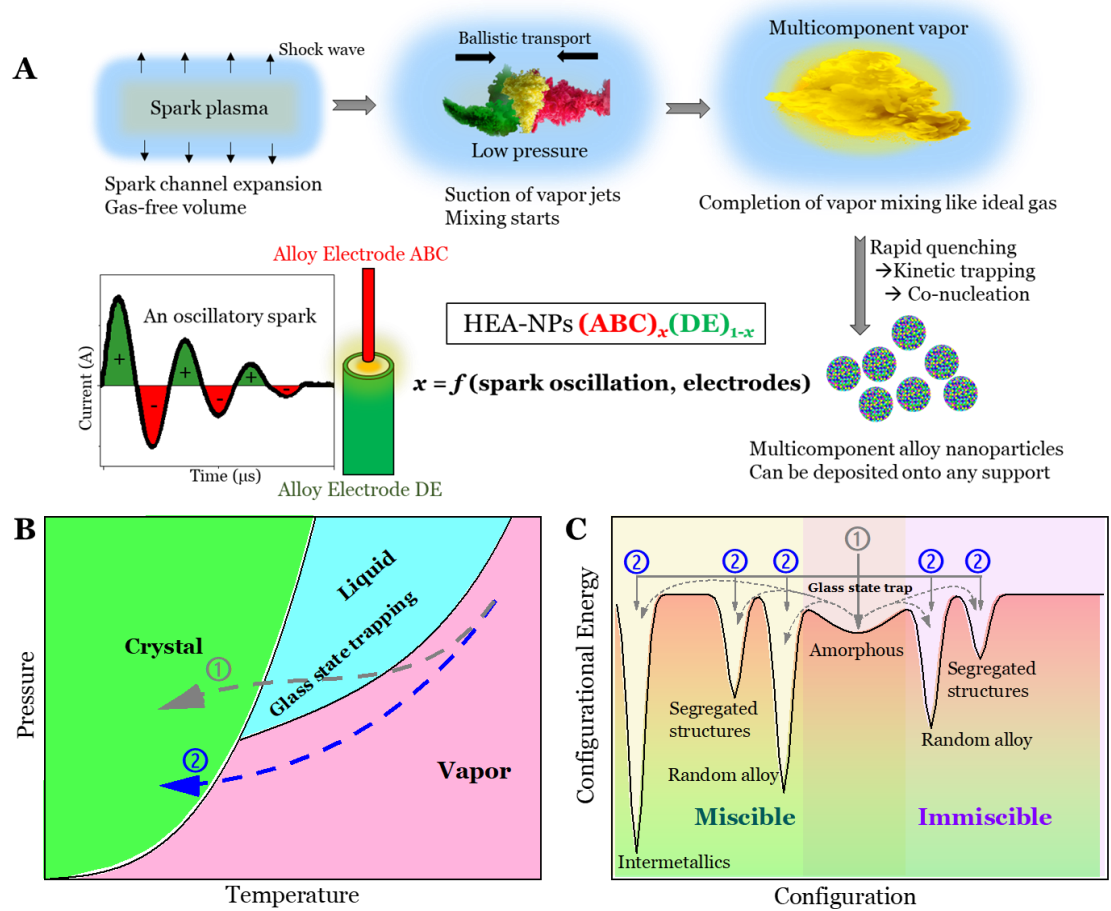

**Fig. 1. Spark mixing principle underlying alloy nanoparticles (NPs).** (**A**) Schematic of the spark mixing mechanism. A spark plasma channel starts to expand and repels the surrounding gas, thus creating a low-pressure region, which draws the vapours injected from different electrodes upon oscillatory sparking. The vapour jets experience ballistic transport toward each other (ca. > $10^3$ m/s) and complete ideal mixing at a high temperature and low pressure. The mixed vapours then co-nucleate and condense into alloy NPs. (**B**) Schematic phase diagram. Cooling path 1 passes through a liquid phase, and allows trapping the configuration into an amorphous/glass state (kinetically frozen liquids). Cooling path 2 (in our NP production) shortcuts the vapour transformation to form crystals. (**C**) Energy landscape of various NP forms in miscible (light-yellow area) and immiscible (light-purple area) systems, overlapping in an amorphous state. For kinetically trapped NPs, the dashed lines for path 1 illustrate possible recrystallizations from an amorphous state. Path 2 leads directly to corresponding crystalline states in the vapour–solid transformation.

An oscillatory spark with a duration of microseconds (fig. S2) controls the feeding composition of the vapour jets from a pair of different-material electrodes[21] (red wire and green cylinder in Fig. 1A); the gas dynamics processes can promote complete vapour mixing. The wire-cylinder electrodes (fig. S3, table S3) attain a high flow velocity in the inter-electrode gap, which not only realizes the production of ultrasmall NPs[22] but also enables fast quenching. Both consequences are proven to be beneficial to forming alloy NPs with size-stabilized mixing (see below). Reversing the polarity of the spark discharge causes the electrodes to alternate as the momentary cathode[21]. For example, the first positive half-cycle (filled in green, Fig. 1A) momentarily makes the green cylinder as the cathode, whereas the red wire takes over as the cathode in the second half-cycle (filled in red). The momentary cathode dominates material vaporization owing to cation bombardment (cations have greater mass/energy than free electrons in the spark plasma). The oscillation waveform then sets the ratio of the spark energy deposited into each electrode, thereby controlling the vapour feeding composition[21]. Such an oscillating waveform is tunable by altering the electrical factors (more details are given in section S3). Meanwhile, the expanding spark channel works like a piston[23–25], moving away the buffer gas and decreasing the pressure within the channel in the wake of the shock wave (Fig. 1A). The vapour jets are then drawn into the channel volume and mixed like an ideal gas at high temperature and low pressure[26]. At the early filling stage, the vapour jets move toward one another within 1 μs[27], characterized by vapour thermal velocities of ca. $10^3$ m/s[25,28]. In the mixed vapour, the high kinetic energy of the atoms can force intermixing between immiscible elements (regardless of the magnitude of positive $\Delta H_{mix}$ in bulk forms). The electronic nature of this formed metallic bonding enables almost infinite mixing possibilities. Upon rapid quenching ($10^7$–$10^9$ K/s[29,30], promoted further by the wire-cylinder electrodes[22]), the mixed vapour allows co-nucleation to form mixed nuclei clusters, which are kinetically trapped and subsequently grow into NPs according to the feeding composition (see sections S4, S5 for details). We therefore coin the term "sparking mashups" to describe the powerful mixing process "alloying any dissimilar metals in NPs". In such continuous NP synthesis, we strongly dilute the aerosol to terminate NP growth prior to agglomeration[31]. Adding high-diffusivity hydrogen to an inert gas prevents gas-impurities interactions with NPs, thereby increasing NP crystallinity[32] and suppressing NP oxidation (figs. S4–S6). The resulting ultrasmall NPs (singlets[31]) are then captured onto any type of support for further immobilization, ready for desired applications (e.g., 3D nanostructures, catalysts).

Generally, alloy formation follows a kinetic pathway that leads metallic ingredients into a thermodynamically stable state. According to pressure-temperature phase diagrams, the thermal history of NP growth can follow two paths from the vapour to the crystal state (Fig. 1B). Path 1 goes through a liquid (i.e., glass-like) state that enables trapping the amorphous state, whereas along path 2, the vapours always directly transform into crystals (given that no glass-forming agents are added). Thus, path 2 offers the more efficient pathway to any desired alloy. The topology of the phase diagram (Fig. 1B) is generally preserved for metal vapours condensing at sufficiently low partial pressure[33]. The sparking mashups (as a vapour source[31]) here is assumed to follow path 2 via rapid quenching to form crystalline NPs of random alloys. In Fig. 1C, we also schematically draw the energy landscape to illustrate different states of the NPs in the form of intermetallics, random alloys, as well as segregated and amorphous structures, while mimicking the energy magnitudes in immiscible and miscible systems. Because intermetallics are excluded from immiscible systems, kinetically suppressed segregation leads the random alloys to become thermodynamically stable at the nanoscale. Relieving the kinetic trapping, however, facilitates phase segregation[2,19]. In miscible systems, random alloys can become metastable because of the presence of intermetallics[34], which are more stable, but the random alloys in our process are kinetically trapped and then stabilized by the nano-size effect, as discussed in detail below.

**Nanosize-stabilized mixing**

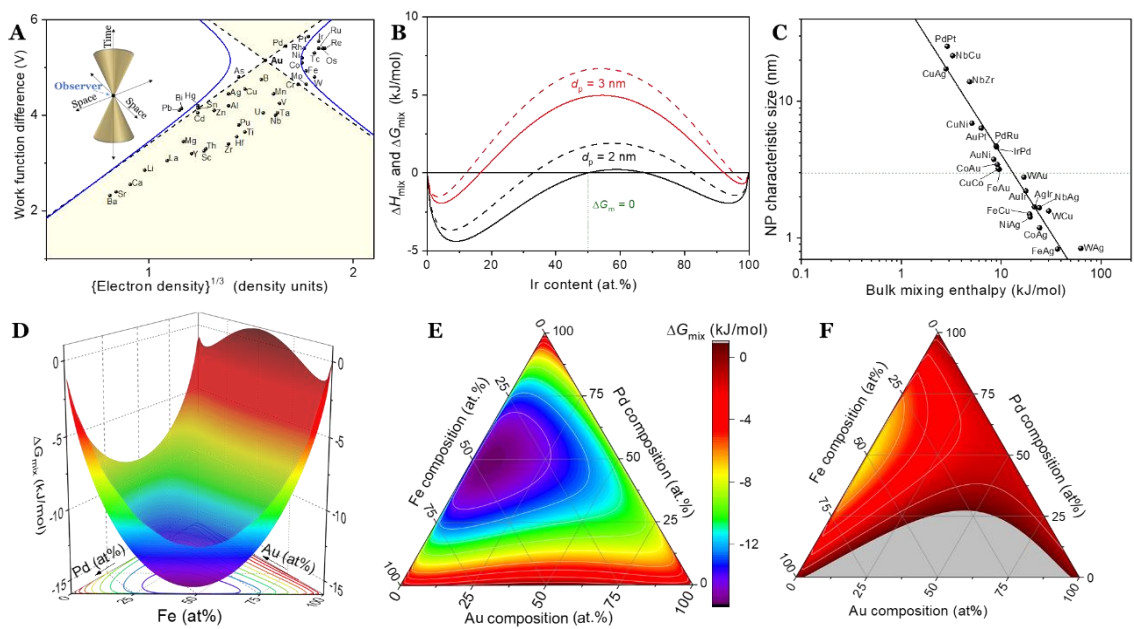

**Fig. 2. Nanosize effect on immiscible systems.** (**A**) Miedema plot (dashed lines) for pairwise bulk miscibility (adapted from ref.[35]) while accumulating the nanosize effect (blue curves). Elements crossing the lines and curves represent equal enthalpy to the Au "observer" of the light cone (inset). Elements inside the yellow area are miscible with Au in the bulk ($\Delta H_{mix}<0$) but those outside are not ($\Delta H_{mix}>0$). The blue hyperbola highlights the nanosize effect, which engulfs some elements that are bulk immiscible with Au. (**B**) $\Delta H_{mix}$ (dashed lines) and $\Delta G_{mix}$ (solid lines) of the 2 and 3 nm IrAu alloy NPs (black and red, respectively) while showing a shift toward more negative values than the bulk (>0). (**C**) Characteristic sizes of a series of immiscible binary alloy NPs. Paraboloid surface (**D**) and contour plots (**E**) of $\Delta G_{mix}$ of ternary alloy NPs (AuPdFe). (**F**) Contour plot of $\Delta G_{mix}$ for the bulk AuPdFe system. (D)–(F) share the same colour bar and refer to 3-nm NPs.

The Miedema theory[35] enables the pairwise classification of miscible and immiscible bulks (fig. S1) (for details, see section S7) and guides the design of HEAs. The electron density at the boundary of the Wigner–Seitz cell and work function describe the interface phenomena on the atomic scale. As a result, the electron density or work function can be used to estimate the enthalpy effects when bringing dissimilar elements into contact (treating atoms as building blocks in metallic states). As shown in Fig. 2A, the dashed lines draw a shape like a light cone (filled in yellow; inset) with Au as the "observer." Such an "observer" can be changed to any element by translating the crosshair (dashed lines) accordingly. The elements touching the borders have an enthalpy equal to that of Au ($\Delta H_{mix} \approx 0$), whereas the ones inside (outside) are miscible (immiscible) with Au in the bulk. Interestingly, currently known HEAs mainly involve the elements along the borders with any desired "observer" ($\Delta H_{mix} \approx 0$). This is because the electron density redistribution at the cell borders compensates the work function adjustment, and substitution entropy stabilizes the mixing phase, which is the essence of HEAs[1].

Considering the surface alloying of immiscible metals[36], the nanosize effect also manifests the role of the surface in mixing. Atoms on a curved NP surface have fewer neighbours and bonds than those on a flat surface, thereby decreasing the surface/strain/free energy. This energy-decrease enforces surface mixing, which propagates inside the NPs. Framing structures (e.g., twin boundaries) erect energy barriers against segregation and call for a spatially variable short-range "order parameter"[37]. The gradient of this parameter describes the mixing depth, which correlates with the characteristic NP size. To understand this nanosize effect, we develop a thermodynamics approach (section S8) for distinguishing the nanoscale stability of alloys from general miscibility trends. The cohesion energy of size-restricted NPs yields an additional negative term in $\Delta H_{mix}$. This term reconstructs the separatrix straight lines that form the blue hyperbolae; thus, the enclosed area identifies an updated miscibility (e.g., Ni, Co) with Au in NPs. To further explain this nanosize effect, we also show a negative shift in both $\Delta H_{mix}$ and $\Delta G_{mix}$ for different sizes of AuIr NPs (Fig. 2B; 24 other types of immiscible binary alloy NPs in fig. S8), where more negative values are shown for the case of 2 nm than those for 3 nm. The equimolar ratio used here (green dashed line) generally requires a maximum shift in $\Delta G_{mix}$ to zero, thereby requiring the strongest nanosize effect. We therefore define a characteristic size (eq. S17), below which the immiscible systems become thermodynamically stable at any molar ratio. The characteristic sizes suffer a linear decline with increasing $\Delta H_{mix}$ for a series of immiscible systems (Fig. 2C), and the green dashed line marks the size of most of the NPs produced here. Hence, a smaller NP is required to compensate a greater positive enthalpy and brings $\Delta G_{mix}$ to zero at an equimolar ratio or to a negative value at other ratios (fig. S9). Such a nanosize effect can be quantified by an entropy-like term in $\Delta G_{mix}$ (Eq.1 below).

Similarly, the nanosize effect also works in ternary alloy NPs; their characteristic sizes become larger than those in binary systems because of the increased entropy. Plotting $\Delta G_m$ of AuPdFe NPs (ca. 3 nm) as a function of each of the three components forms a paraboloid surface, which enables qualitatively comparing the elemental contents (Fig. 2D). Pairwise $\Delta G_{mix}$ magnitudes are reflected by the depth of the curves projected on each plane, ranked in descending order in magnitudes as Pd–Fe, Pd–Au, and Au–Fe. The most stable state ($\Delta G_{mix}$ = −15.8 kJ/mol) has the highest amount of Pd (Pd:Fe:Au = 49:44:7). This asymmetry is also demonstrated in the contour plots (Fig. 2E, and 16 other types of ternary alloy NPs in fig. S7); however, the symmetry attempts to restore itself at low mixing enthalpies and/or high mixing entropies (i.e., multiple elements). Notably, AuPdFe NPs of nearly any composition are stable at room temperature because of their negative $\Delta G_{mix}$ in almost all regions of the phase diagram (Fig. 2E). Such a broad composition range suggests that potential HEAs may have been overlooked, because only a few discrete compositions are typically attempted using conventional methods. Without the nanosize effect, the corresponding bulk materials cannot form as many stable alloys at room temperature (Fig. 2F, and 16 other types of bulk systems in fig. S10).

To extend to MA-NPs containing $n$ elements ($n \geq 2$), the size-bearing term contributes a negative value to $\Delta H_{mix}$ (section S8) while following the form of $\Delta S_{mix}$ with an effective temperature $T_M$. This size-bearing term is then added to the original entropy in HEAs, expressed by the mixing free energy $\Delta G_M^{(S)}$:

$$\Delta G_M^{(S)} = R(T_M + T) \sum_{i=1}^{n} (x_i \ln(x_i)) \qquad (1)$$

where $R$ is the gas constant, $x_i$ denotes an atomic composition of element $i$, and $T_M$ is inversely proportional to the NP size (eq. S20). Intriguingly, such a size effect on entropy reveals a close similarity to the relation between the Hawking temperature and the size of a black hole, which is a known perfect mixer. Even if drastic unmixing occurs in the bulk alloy at $T = 0$ K [7], $T_M$ still enforces the entropic mixing in MA-NPs. To quantify its contribution, we estimated that 3-nm NPs, even for binary alloys, already have entropy comparable with that of a bulk HEA containing at least 25 elements (section S8). This higher entropy explains the successful synthesis of large HEA-NPs[2,38], but in immiscible binary alloy NPs, smaller NPs are required to avoid phase segregation[2].

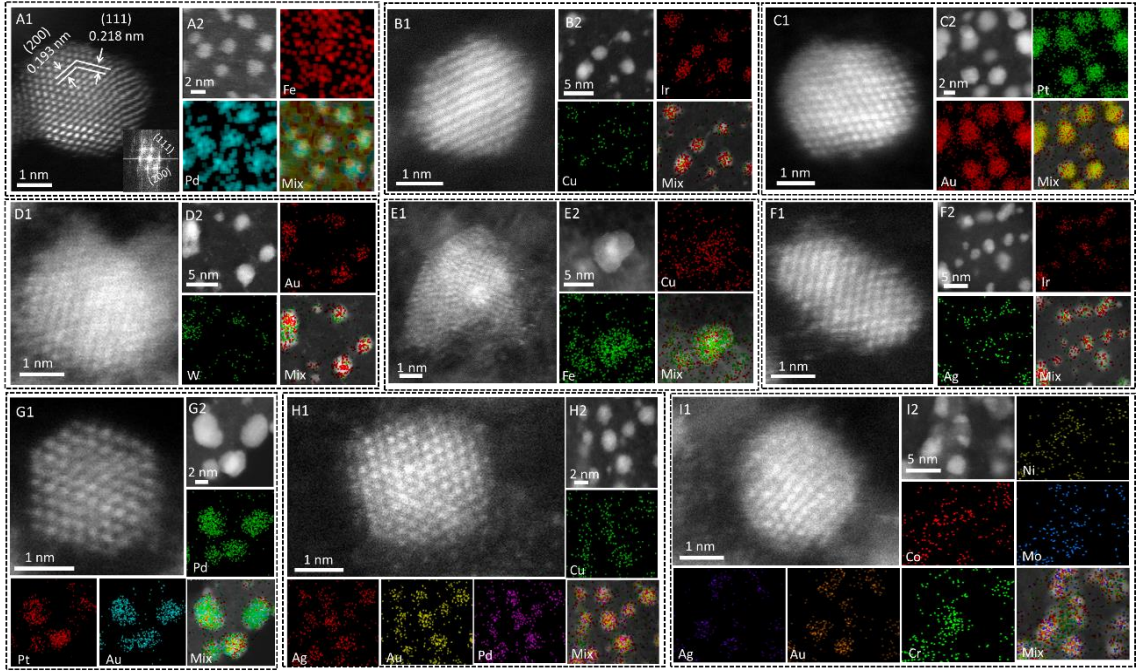

**Fig. 3. Elemental characterization of MA-NPs.** High-angle annular dark field–scanning transmission electron microscopy (HAADF-STEM) images and elemental maps of MA-NPs. (**A**)–(**F**) Binary miscible (FePd) and immiscible (IrCu, PtAu, AuW, CuFe, IrAg) alloy NPs. (**G**)-(**I**) Ternary (AuPdPt), quaternary (AuAgCuPd), and senary (NiCoMoAgAuCr) alloy NPs.

Sparking mashups enables vapour mixing and subsequent co-nucleation to form kinetically trapped mixed clusters, which in turn form alloy NPs with size-stabilized mixing. The small NPs (<5 nm) were achieved by strongly diluting the aerosol prior to agglomeration[31] with the help of wire-cylinder electrode-configuration and then immobilizing the NPs onto a substrate by various means of aerosol deposition (e.g., filtration, electrical attraction). We then used aberration-corrected scanning transmission electron microscopy (Cs-STEM) to further ascertain the crystal structure and identify the compositional uniformity (Fig. 3). As representative examples for this analysis, we carefully selected NPs from among binary immiscible (IrCu, PtAu, WAu, CuFe, IrAg, WAg, and IrAu), and binary miscible (FePd, PtW, and ZrAu), ternary (AuPdPt, and AuCuPd), quaternary (AuAgCuPd and AuNiCuPd), quinary (NiCrFeAuAg and NiCrCoAuAg), and senary (NiCrCoMoAuAg) alloy NPs (Fig. 3, fig. S11).

Cs-STEM energy-dispersive X-ray spectroscopy (EDX) maps show the compositional uniformity of all these NPs (Fig. 3, fig. S11). Compositional analysis results at the microscale are more or less consistent with those at the macroscale (inductively coupled plasma mass spectrometer (ICP-MS), figs. S12, S13). The speckling effect on binary alloy NPs serves as the supplementary evidence of the well-mixed states[12]. STEM–EDX compositional analysis indicates that the NPs have a similar composition, irrespective of their size (below the threshold, Fig. 2C) and immiscibility (fig. S14). From the fast Fourier transform (FFT), $d$-spacings of $0.218 \pm 0.02$ and $0.193 \pm 0.001$ nm are obtained, corresponding to (111) and (200) interplanar distances of a face-cantered cubic (fcc) structure (Fig. 3A), respectively, with lattice parameter $a = 0.3818 \pm 0.03$ nm, which is closer to the parameter of Pd ($a = 0.3908$ nm) than to that of Fe ($a = 0.2866$ nm). We also show the ultrasmall sizes and the good dispersity of the produced MA-NPs (figs. S15–S23), regardless of the elemental combinations (figs. S24–S28). Powder X-ray diffractometry (XRD) and X-ray photoelectron spectroscopy (XPS) further confirm the alloy nature of a large amount of NPs (figs. S29–S33). In Cs-STEM imaging, we observed that many of the particles present multiple twins (fig. S34). For twin formation, the Laplace pressure of NPs pushes different parts of the crystalline NP against each other, thus causing plastic deformations and sliding motion. The latter produces mixing and twinning planes to frame the crystal structure, further explaining why such fast cooling leads to, for example, crystalline instead of amorphous NPs. This case suggests that the vapour–solid time–temperature–transformation (TTT) curve is not shaped like a standard C-curve but forms a low-temperature "crystalline tail," similar to a martensite transition tail in steels.

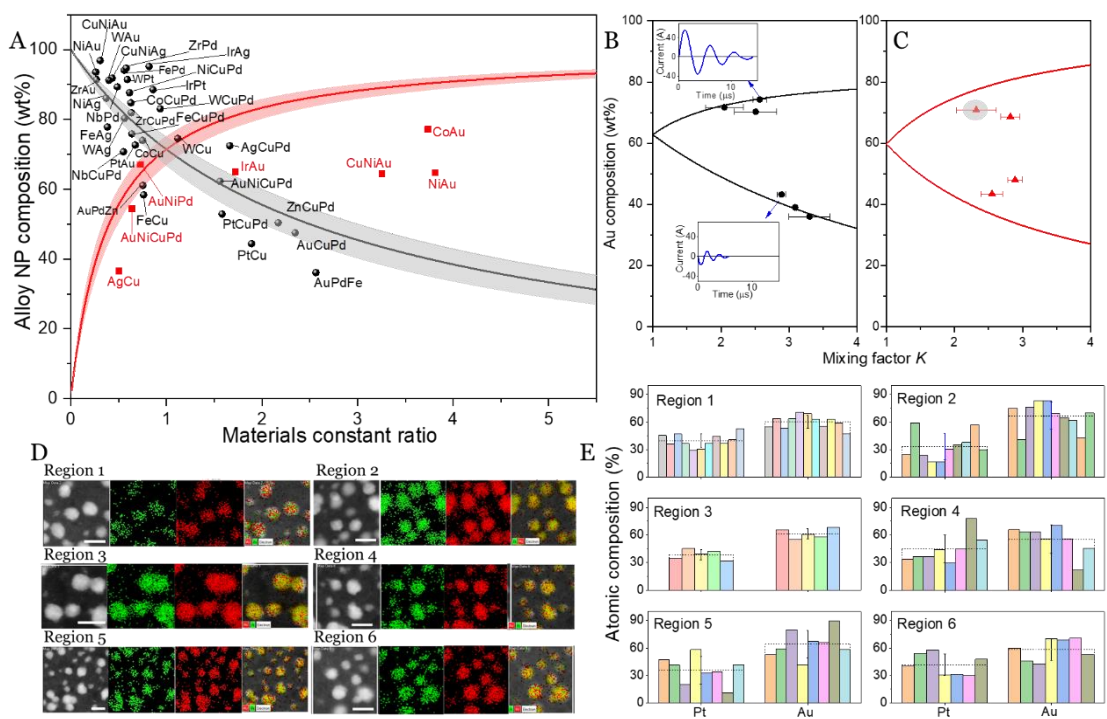

**Fig. 4. Alloy NP compositional control.** (**A**) Composition of various alloy NPs (wt%) with mixing factor $K$. The compositions (wt%) here always adhere to the material of the cylindrical electrode (table S1). Black and red curves ($K = 2.5$) from model predictions correspond to the positive and negative polarity of the current source used in sparking mashups, and the shadows

indicate the ±0.5 variation in $K$. The compositions of NPs presented in scatters were determined by ICP-MS. (**B**), (**C**) Dependence of the composition on $K$ for ZrAu and PtAu alloy NPs, respectively. The model predictions (curves) match with the measurements (scatters) from ICP-MS. The insets show the current oscillation at different $K$ values (indicated by arrows). The error bars indicate the standard deviation of six current oscillation measurements taken at different times. (**D**) STEM images and elemental maps of the PtAu NPs in different regions (scale bars 5 nm; red and green represent Au and Pt, respectively) and (**E**) their compositional (at%) uniformity. Each bar represents the elemental composition of a NP in the given region, and error bars indicate the standard deviation of the number of measured NPs. The PtAu NPs were produced with a $K$ value marked by a grey circle in (C).

Next, we demonstrate the compositional controllability and uniformity of a wide range of alloy NPs. Derivation of the mixing factor $K$ and materials constant ratios is discussed in section S3 (table S4). Scatters represent the NP weight percent (for the sake of unity between the model prediction and ICP-MS measurements) of the element(s) from the cylindrical electrode (Fig. 4A), which was made of alloys or unary metals (fig. S3, table S3). For example, we produced AuNiCuPd alloy NPs by using a pair of alloy electrodes consisting of a AuNi wire and a CuPd cylinder. The weight percent in Fig. 4A thus denotes the summed contents of Cu and Pd. Using a fixed $K$ but changing only the electrode materials, the data presented in black show the composition of various NPs generated via sparking mashups that was powered by a positive current source, whereas a negative polarity was used to synthesize the NPs with compositions presented in red (Fig. 4A). Switching only the polarity of the current source results in a change in the NP compositions (e.g., NiAu, IrAu, AuNiCuPd, CuNiAu; fig. S35, S36). The shadows indicate the influence of $K$ variations (2.5 ± 0.5) on the NP compositions in model predictions. The measured compositions largely fall into the regions of the model predictions. The maximum discrepancies are 10–20 wt% for certain alloys (e.g., CuNiAu, NiAu), which is mild when considering model simplifications and $K$ variations among many sparks (ca. 300 000).

To further show the ability to control the composition of a specific type of alloy NPs, we altered $K$ through the capacitance (or other parameters, as explained in section S3, fig. S37). Fig. 4B,C shows the compositional consistency between the model predictions (curves) and the ICP-MS measurements (symbols) for ZrAu and PtAu NPs, which represent miscible and immiscible binary systems, respectively. The slight discrepancy (ca. 10% for PtAu) can be attributed to material transfer between electrodes (fig. S38). The insets show the difference in the amplitude/frequency of the spark oscillation at different $K$ values, which control the NP composition. To analyse the composition, we choose the PtAu alloy NPs (marked by a grey circle in Fig. 4C) as a powerful demonstration of our alloying strategy, because their immiscibility raises challenges in conventional syntheses. The compositional uniformity for each element is confirmed by statistically analysing the EDX data from six different regions (Fig. 4D and E), across which the Au composition varies by approximately 10%, lower than the > 40% variation reported previously[12,39]. Thus, the immiscibility therefore proves not to be problematic for NP composition uniformity, validating the powerful sparking mashups to from alloy NPs with nanosize-stabilized mixing (i.e., complete vapour-mixing and subsequent kinetic trapping). The average composition among five different areas varies by approximately 3 at% (fig. S12). The Au composition (67 ± 3 at%, fig. S12) according to the EDX analysis of these areas exhibits an approximately 4 at% discrepancy relative to that of ICP-MS (71 at%). We also evaluated the compositional variation for other MA-NP samples (figs. S12, S13), and the results were also largely similar.

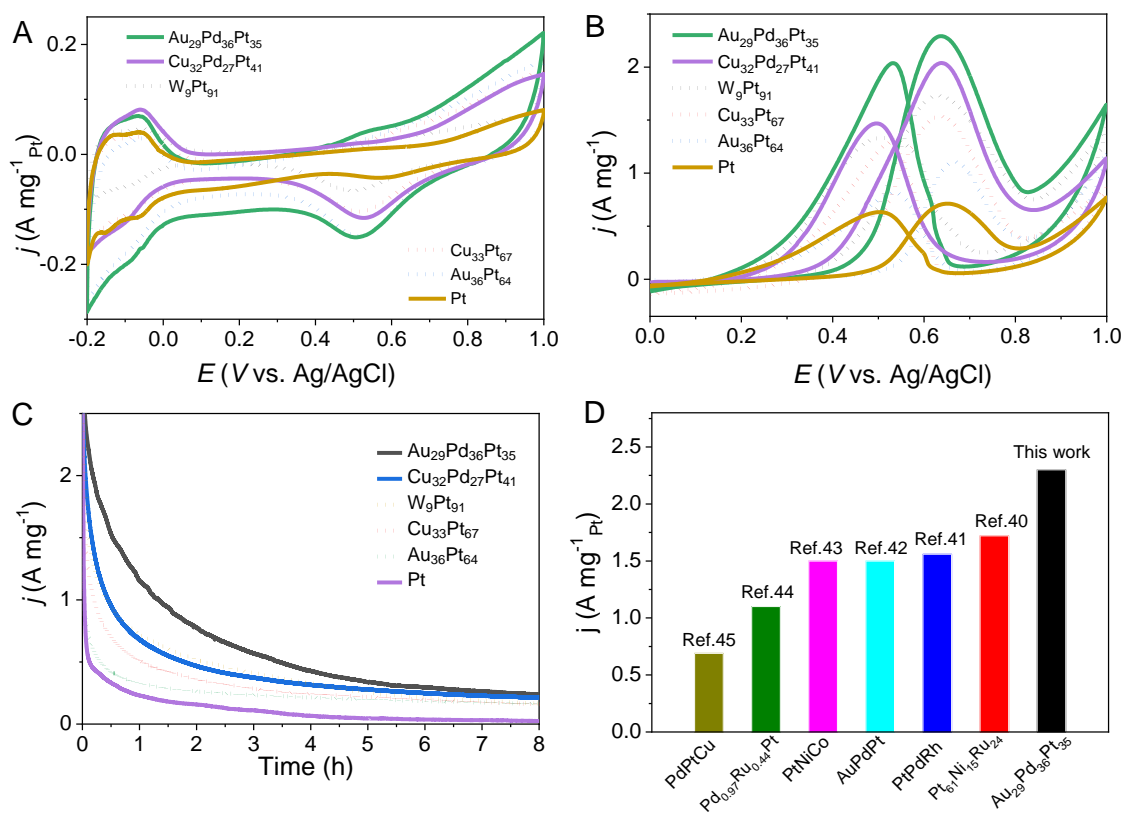

**Fig. 5. Catalytic performance of the alloy NPs.** (A) Cyclic voltammograms (CVs) in 0.1 M $HClO_4$ at a scan rate of 50 mV s$^{-1}$ for $Au_{29}Pd_{36}Pt_{35}$, $Cu_{32}Pd_{27}Pt_{41}$, $W_9Pt_{91}$, $Cu_{33}Pt_{67}$, $Au_{36}Pt_{64}$, and unary Pt NPs. (B) CVs in 1 M methanol and 0.1 M $HClO_4$ at a scan rate of 20 mV s$^{-1}$. (C) Chronoamperometry at 0.6 V at room temperature. (D) Comparison of the mass activity for methanol oxidation with those of previously reported ternary nanocatalysts[40–45].

The alloy NPs produced in this work (e.g., those containing Pt and Pd) are valuable and important electrocatalysts. Combining the alloying effect with ultrasmall, high-purity NPs offered by sparking mashups proves beneficial to electrocatalytic reactions. The Pt and Pd alloy NPs (supported on carbon paper, fig. S39) were used as catalysts for the room-temperature methanol oxidation reaction (MOR), ethanol oxidation reaction (EOR), and formic acid oxidation reaction (FAOR). Here, we take MOR as a model reaction for assessing the catalytic performance of Pt alloy NPs. Integration of the hydrogen absorption–desorption region in the cyclic voltammograms (CVs) (Fig. 5A), followed by Pt mass normalization, yields electrochemically active surface areas (ECSAs) of 121.6 m$^2$ g$^{-1}$ for $Au_{29}Pd_{36}Pt_{35}$, 111.5 m$^2$ g$^{-1}$ for $Cu_{32}Pd_{27}Pt_{41}$, 90.3 m$^2$ g$^{-1}$ for $Au_{36}Pt_{64}$, 76 m$^2$ g$^{-1}$ for $Cu_{33}Pt_{67}$, 80.1 m$^2$ g$^{-1}$ for $W_9Pt_{91}$, and 85.8 m$^2$ g$^{-1}$ for Pt NPs, all of which are considerably larger than those of similar catalysts produced using wet-chemistry methods[46]. We attribute these larger ECSAs to the increased number of surface reactive sites owing to the ultrasmall NPs and their clean surfaces. Meanwhile, variations among these ECSAs are most likely due to metal alloying and slight differences (<

1 nm) in NP size. For all the NPs tested here, methanol oxidation commences at ca. 0.32 V and fully develops to a peak at ca. 0.66 V (Fig. 5B). A comparison of the current densities (table S5) suggests that the Pt alloy NPs show greater mass activities than unary Pt, indicating the positive alloying effect. In particular, $Au_{29}Pd_{36}Pt_{35}$ NPs achieve a peak value of 2.3 A mg$^{-1}$, comparable with the state-of-the-art values (table S6) but higher than those of the ternary nanocatalysts (Fig. 5D). In addition to the mass activities, we show their higher specific activities (peak current densities normalized by ECSAs), benchmarked by the unary Pt NPs (fig. S40). In principle, alloying Pt with other metals can modify the electronic structure, presenting a negative shift in the $d$-band centre relative to that of Pt (fig. S41). Such modifications weaken the binding energy between CO-like intermediates and Pt sites [47], confirmed by a negative shift in the CO stripping peaks (fig. S42). Citing the same reasons, chronoamperometry shows that all the alloy NPs are clearly more stable than unary Pt NPs, and among the alloy NPs, the $Au_{29}Pd_{36}Pt_{35}$ NPs exhibit the slowest decay in mass activity at 0.6 V for 8 h (Fig. 6C). These studies reveal that the alloy NPs exhibit not only increased catalytic activity but also improved durability.

The Pd alloy NPs produced here also outperform unary Pd NPs and similar catalysts in many previous works (tables S7, S8) in catalytic activities for both the EOR and FAOR (figs. S43–S48). Among the Pd alloy NPs, the $Au_{29}Pd_{36}Pt_{35}$ NPs again achieve a high mass activity of 4.9 A mg$^{-1}$ for the EOR. Further, $Ir_{20}Pd_{80}$ (a bulk-immiscible pair) and $Au_{23}Fe_{17}Pd_{60}$ NPs show a mass activity of 1.5 A mg$^{-1}$ for the FAOR and 2.9 A mg$^{-1}$ for the EOR, respectively. Interestingly, many of the alloys produced here have never been reported to date, either as nanomaterials or bulk materials (table S1), and those reported have rarely been studied as electrocatalysts (e.g., WPt, IrPd, and AuFePd). However, we successfully identified their excellent catalytic performance, such as the greater CO tolerance of $W_9Pt_{91}$ NPs than that of the other catalysts (fig.S42). Although such alloy NPs have not yet been optimized, and the mechanisms underlying their catalytic activities are yet to be explored, their unique performance offers the potential for developing novel electrocatalysts.

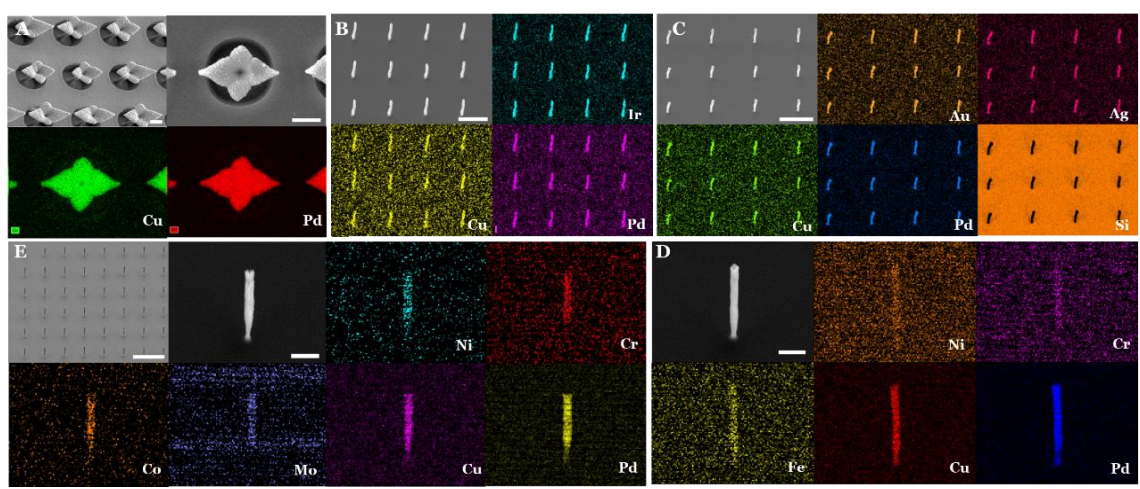

**Fig. 6. 3D-printed alloy nanostructures.** (**A**) Binary alloy (CuPd) nanostructure arrays. The SEM image (tilted view) shows the array of 3D flower-like structures. (**B**) Ternary alloy (IrCuPd) nanopillar arrays. (**C**) Quaternary alloy (AuAgCuPd) nanopillars. The EDX signal of Si is from the substrate used in 3D nanoprinting. (**D**) Quinary alloy (NiCrFeCuPd) nanopillars. (**E**) Senary alloy (NiCrCoMoCuPd) nanopillars. (A), (D), and (E) have scale bars of 1 μm, and the SEM image in (E) has a 20-μm scale bar, whereas both (B) and (C) have scale bars of 10 μm. More detailed SEM images of the nanostructures are presented in figs. S49–S52.

Importantly, sparking mashups can also be integrated into additive manufacturing, commonly known as 3D printing, so that new structural properties can be offered on top of the positive alloying effect (Fig. 5). Ongoing research aims to downscale metal 3D printing to the nanoscale and provide flexible material choices. Such developments, however, are severely limited by particle technologies and printing techniques. Here, we fulfil this aim by implementing sparking mashups as a particle source for a recently developed 3D nanoprinting system[48]. In this 3D printing, patterned dielectrics help to hold charges as fundamental dipole components to reshape electric field lines into a virtual/noncontact nozzle without downscaling restrictions. Charged NPs directly produced by sparking mashups are then precisely guided through these virtual nozzles to a surface, where 3D nanostructures are printed in parallel[49,50]. The resulting ordering energy is estimated to be at least two orders of magnitude larger than the disordering/kinetic energy due to Brownian motion at room temperature[49]. Therefore, this relationship, together with the fast kinetics of aerosol technology, enables a general principle for NP assembly in gases rather than in liquids ( note that relevant diffusion coefficients in gases are three orders of magnitude higher than those encountered in liquids). Such a valuable result compensates for some of the deficiencies in colloidal chemistry, which have largely dominated the field of NP assembly to date but poses serious challenges to purity, HEA fabrication, material limitation, and printing speed. We then harness this system to develop hierarchical nanostructures with single-NP precision with high purity. SEM-EDX mapping (Fig. 6) confirms that the resulting nanostructures consist of binary (FePd), ternary (IrCuPd), quaternary (AuAgCuPd), quinary (NiCrFeCuPd), and senary (NiCrCoMoCuPd) alloy components. The alloy NPs produced here are used as building blocks, and the EDX maps suggest that the 3D nanostructures retain the alloy phase. Note that the 3D structures were printed on a Si substrate, leading to the corresponding EDX signal (shown in dark yellow, Fig. 6C). Only 20 min was required to print approximately 3000 IrCuPd nanopillars that have nearly identical geometries with a diameter of approximately 650 nm and a length of 8.5 μm. We thus estimate a printing speed of 25 μm/s, faster than that of most metal 3D printing with similar feature sizes[51]. To demonstrate the flexible choice of substrates, NiCrCoMoCuPd nanostructures with a width of 80 nm and a height of 300 nm were also printed on indium tin oxide (ITO) glass (figs. S50–S52). Until now, HEA structures have never been downscaled to such small feature sizes or materialized over such a wide range of material compositions. The results here thus enable the development of 3D printing for new HEA nanostructures with sufficient mechanical robustness for actual devices[11].

**Remarks and discussions**

To ensure the thermal stability of the NPs, we strongly dilute the aerosol to terminate NP growth in the gas phase before agglomeration[31], and we capture these ultrasmall NPs (< 5 nm, figs. S15–S23) onto a substrate for further immobilization. Because of the liquid-free nature of this NP synthesis, the resulting pure surface facilitates NP agglomeration upon deposition. However, this agglomeration is insufficient to lead to the phase segregation observed in immiscible systems[52] because of the size-stabilized mixing (Fig. 2C). Continuous deposition covers the substrate either with a particulate film or an array of NP structures. The thermal stability is still preserved, irrespective of the film thickness and height of the structure. The mixing state should also not be influenced as long as nanograin boundaries are maintained. On the other hand, the random dispersion of dissimilar metal atoms leads to variations in the heights of the activation energy barrier for hopping diffusion. This barrier height distribution can be described phenomenologically by the same size-dependent term, $T_M$ (Eq. 1, section S8), because the surface terms mimic the entropy contributions to free energy (eq. S19, S20). Therefore, the random and quenched disorder in MA-NPs can exhibit unusual stability with respect to surface diffusion being suppressed by non-Arrhenius terms (eq. S21). This nanosize effect on thermal stability fundamentally differs from the traditional "smaller is less stable" view[9]. Such important information can aid the development of stable HEA nanostructured alloys (films or 3D structures). Sparking mashups can be scaled up to either substrate-based products or continuous micro/nano manufacturing, and this scalability was already realized[53].

**Conclusions**

Herein, we have introduced the use of sparking mashups to mix 55 types of alloys in ultrasmall NPs, ranging in composition from binary to senary alloys, with compositional controllability and uniformity. Remarkably, we even break the miscibility limits to successfully mix bulk-immiscible elements in alloy NPs. Two electrodes of different materials are alternatingly vaporized upon oscillatory sparks lasting several microseconds. These oscillatory sparks act like a piston to enforce vapour mixing while controlling the feeding composition. The resulting vapour allows the co-nucleation of mixed clusters, which are kinetically trapped and subsequently grow into NPs of random alloys in accordance with the feeding composition. In addition, we develop a thermodynamic approach to explain the size-stabilized mixing, which mimics the role of mixing entropy in HEAs. The alloy NPs are demonstrated for fuel-cell catalysis and 3D printing nanostructural arrays. Our general approach to sparking mashups of alloy NPs containing any dissimilar metals opens the way to new, uncharted territories, wherein many novel materials, structures, theories, and applications are waiting to be discovered. Further, the results of this study can help overcome challenges in unconventional alloy synthesis and explore new frontiers in HEAs, catalysis, and 3D nanoprinting.

**Data availability** The data that support the findings of this study are available from the authors on reasonable request, see author contributions for specific data sets.

**Acknowledgments** We thank Prof. S. Cho for the Cs-STEM measurements and Prof. Schmidt-Ott for the discussions. This work was supported by the Global Frontier R&D Program for Multiscale Energy Systems (2012M3A6A7054855) from the National

**Author contributions** J.F. and M.C. conceived the idea, analysed the data, and wrote the manuscript. J.F., P.P., and D.C. prepared the supplementary material. P.P. conducted the theoretical work, and J.F. and P.P. analysed and wrote the theoretical parts. J.F. designed and carried out the experiments; some experiments were also performed by Y-H.J. The results for catalysts and TEM were obtained by D.C. and J.Y. M.C led the work. All authors discussed the contents of this work.

**Competing interests** The authors declare no competing interests